\documentclass[12pt]{article} 
\usepackage{amsmath}
\newcommand{\be}{ \begin{equation}}
\newcommand{\ee}{\end{equation}} 

\begin{document} 
\def\theequation{\arabic{section}.\arabic{equation}} 
\begin{titlepage} 
\title{Lema\^itre model and cosmic mass} 
\author{Valerio Faraoni\\ \\ 
{\small \it Physics Department and  STAR Research 
Cluster, Bishop's University}\\ 
{\small \it 2600 College Street, Sherbrooke, Qu\'ebec Canada 
J1M~1Z7}\\
{\small \it tel + 819 822 9600 ext.~2490, fax 
+~819~822~9611, email~vfaraoni@ubishops.ca}
}
\date{} \maketitle 
\thispagestyle{empty} 
\vspace*{1truecm} 
\begin{abstract} 
The mass of a sphere of simmetry in the Lema\^itre universe is 
discussed using the Hawking-Hayward quasi-local energy and 
clarifying existing ambiguities. A covariantly conserved current 
introduced by Cahill and McVittie is shown to be a multiple of the 
Kodama energy current.
\end{abstract} 
\vspace*{1truecm} 
\begin{center} 
{\bf Keywords:} inhomogeneous cosmologies; cosmic mass; 
cosmic parameters. 
\end{center} 
\end{titlepage}

\def\theequation{\arabic{section}.\arabic{equation}}

\section{Introduction}
\label{sec:1}
\setcounter{equation}{0}

The Lema\^itre model \cite{Lemaitre} is a spherically symmetric 
inhomogeneous 
universe which solves the Einstein equations and generalizes the 
better known Lema\^itre-Tolman-Bondi (LTB) \cite{Tolman, 
Bondi} geometry to the case of  
non-zero pressure. Inhomogeneous universes (see 
Ref.~\cite{Krasinskibook} for a comprehensive review) have been the 
subject of much work in recent years because of the attempts to 
explain the current acceleration of the universe without an {\em ad 
hoc} dark energy or cosmological constant and without abandoning 
general relativity (see \cite{Bolejko} for a review). This work on 
LTB models has revived interest also in Lema\^itre models and 
inhomogeneous cosmologies in general. The issue has been raised 
recently 
of what is the physical mass contained in a sphere in the 
Lema\^itre universe \cite{AH}. Over the years, there have been 
various proposals (reviewed in \cite{AH}). More recently, LTB 
models have been studied in various contexts \cite{recent} and the 
study has moved to a higher degree of sophistication with the 
introduction of perturbations of LTB models 
\cite{LeithesMalik2014}; similar to the case of 
Friedmann-Lema\^itre-Robertson-Walker (FLRW) spacetimes, 
perturbations allow research on more sophisticated physics and open 
the door to the use of tools such as the temperature fluctuations 
in the cosmic microwave background \cite{Durrerbook}. As a 
starting point for this and future developments, it would be 
reassuring to know that we understand the physics of the {\em 
unperturbed} 
Lema\^itre universe, including its energy. Here we show that 
approaching this issue using the Hawking-Hayward quasi-local energy 
\cite{Hawking, Hayward1} and the Kodama vector 
\cite{Kodama} dissipates some ambiguities in the recent 
literature. In the 
presence of spherical symmetry (and, therefore, also in the 
Lema\^itre universe), the Hawking-Hayward quasi-local energy 
reduces \cite{Hayward1, Hayward2} to the Misner-Sharp-Hernandez 
mass 
\cite{MSH}. Spherical 
symmetry allows one to introduce also the Kodama vector 
\cite{Kodama}, which generates a covariantly conserved current and 
has the Misner-Sharp-Hernandez mass as its Noether charge 
\cite{Hayward1}.

In the next section we discuss the Lema\^itre universe and the 
Misner-Sharp-Hernandez mass contained in sphere of 
symmetry,\footnote{By ``sphere of symmetry'' we mean a 
2-dimensional surface which is an orbit of the isometry of the 
spacetime manifold describing spherical symmetry (of course, such 
orbits exist through any point of a spherically symmetric 
spacetime).} showing how this clarifies doubts that still linger on 
in the literature. The 
following section discusses the Kodama vector and shows that a 
conserved current discussed in Refs.~\cite{CahillMcVittie, 
AH} is a multiple of the Kodama current built out of the Einstein 
tensor and the Kodama vector. 

We use units in which the speed of light and Newton's constant are 
unity and we follow the notation of Wald's 
textbook~\cite{Waldbook}.

\section{Lema\^itre geometry and the mass of a sphere of 
symmetry}
\label{sec:2}
\setcounter{equation}{0}

The spherically symmetric  Lema\^itre line element in coordinates 
$\left\{ x^{\mu} \right\}= \left\{ t,r, \theta, \varphi \right\}$ 
comoving with the fluid 
source is \cite{Lemaitre}
\be\label{1}
ds^2=-\mbox{e}^{2\sigma} dt^2 +\mbox{e}^{\lambda} dr^2 +R^2 
d\Omega_{(2)}^2 \,,
\ee
where $\sigma=\sigma (t, r), \lambda=\lambda (t, r)$, and $R(t, 
r)$ is the areal radius, while $d\Omega_{(2)}^2 =d\theta^2 +\sin^2 
\theta \, d\varphi^2$ is the line element on the unit 2-sphere. The 
simplest stress-energy tensor sourcing the Lema\^itre spacetime is 
that of a perfect fluid
\be \label{2}
T_{ab}=\left( P+\rho \right) u_a u_b +Pg_{ab} \,
\ee
where $\rho (t, r)$ and $P(t, r)$ are the energy density and 
pressure of the fluid as perceived by a comoving observer with 
4-velocity $u^c$ with components $u^{\mu}=\left( 
\mbox{e}^{-\sigma}, 0,0,0 \right)$ in comoving coordinates. 
In general, however, the matter source of the Lema\^itre 
metric~(\ref{1}) is not restricted to be a perfect fluid. 

Following \cite{AH}, we assume 
isotropic pressure (which is a restriction on the full Lema\^itre 
spacetime) and we allow for a cosmological constant $\Lambda$. The 
Lema\^itre universe is a solution of the Einstein equations
\be\label{efe}
G_{ab}+\Lambda g_{ab}=8\pi T_{ab} \,,
\ee
where $G_{ab}$ is the Einstein tensor and $g_{ab}$ is the spacetime 
metric.

In order to discuss the Misner-Sharp-Hernandez mass $M_\text{MSH}$ 
and 
the Kodama vector $k^a$, it is convenient to express the line 
element~(\ref{1})  using the areal radius $R=x^{1'}$ instead of 
$r=x^1$ 
because then $M_\text{MSH}$ and $k^a$ assume straightforward 
expressions. Our conclusions, however, are fully 
coordinate-independent and it will be easy to revert to the 
coordinates $\left\{ x^{\mu} \right\}=\left\{ t,r, \theta, \varphi 
\right\}$. Our goal is to 
recast the line element~(\ref{1}) in the form
\be\label{0}
ds^2=-A(T,R) dT^2 +B(T,R) dR^2 +R^2 d\Omega_{(2)}^2 \,.
\ee
Using the differential relation $dR=\dot{R}dt+R'dr $, where an 
overdot and a prime denote differentiation with respect to $t$ and 
$r$, respectively, the line element~(\ref{1}) becomes
\be \label{3}
ds^2 = - \left( \mbox{e}^{2\sigma}- \mbox{e}^{\lambda}\, \frac{ 
\dot{R}^2}{ R'^2} \right) dt^2 +\frac{ \mbox{e}^{\lambda}}{R'^2} 
\, dR^2 -\frac{ 2\dot{R} \, \mbox{e}^{\lambda} }{R'^2}\, dtdR + R^2 
d\Omega_{(2)}^2 \,.
\ee
The cross-term in $dtdR$ is eliminated by introducing a new time 
coordinate $x^{0'}= T$ defined by
\be\label{4}
dT = \frac{1}{F} \left( dt +\beta dR \right) \,,
\ee
where $\beta(t,r)$ is a function to be determined and $F(t,r)$ is 
an integrating factor which must be introduced to guarantee that 
$dT$ is an exact differential. It satisfies the equation
\be\label{5}
\frac{\partial }{\partial r} \left( \frac{1}{F} \right) = 
\frac{\partial }{\partial t} \left( \frac{\beta}{F} \right) \,.
\ee
Upon substitution of $dt=FdT-\beta dR$, the line element~(\ref{3}) 
assumes the form
\begin{eqnarray}
ds^2 &=&  - \left( \mbox{e}^{2\sigma}- \mbox{e}^{\lambda}\, \frac{
\dot{R}^2}{ R'^2} \right) F^2 dT^2 +2F \left[ 
\beta  \left( \mbox{e}^{2\sigma}- \mbox{e}^{\lambda}\, \frac{
\dot{R}^2}{ R'^2} \right)
- \frac{\dot{R}}{ R'^2} \, \mbox{e}^{\lambda} \right] dTdR 
\nonumber\\
&&\nonumber\\
& \, & + \left[  \frac{\mbox{e}^{\lambda}}{ R'^2}  
- \left( \mbox{e}^{2\sigma}- \mbox{e}^{\lambda}\, \frac{
\dot{R}^2}{ R'^2} \right) \beta^2
+2\beta  \frac{\dot{R}}{ R'^2} \, \mbox{e}^{\lambda}\right] dR^2 
+ R^2 d\Omega_{(2)}^2 \,. \label{6}
\end{eqnarray}
Setting
\be\label{7}
\beta(t,r)= \frac{ \dot{R}\, \mbox{e}^{\lambda} }{  R'^2 \left( 
\mbox{e}^{2\sigma}- \mbox{e}^{\lambda}\, \frac{
\dot{R}^2}{ R'^2} \right)} 
\ee
one obtains
\be
ds^2 =  - \left( \mbox{e}^{2\sigma}- \mbox{e}^{\lambda}\, \frac{
\dot{R}^2}{ R'^2} \right) F^2 dT^2 
+ \frac{  \mbox{e}^{\lambda} \mbox{e}^{2\sigma}}{
R'^2 \left( \mbox{e}^{2\sigma}- \mbox{e}^{\lambda}\, \frac{
\dot{R}^2}{ R'^2} \right)} \,  dR^2 
+ R^2 d\Omega_{(2)}^2 \,, \label{8}
\ee
which is of the form~(\ref{0}) with
\begin{eqnarray}
A(T,R) &=& \left( \mbox{e}^{2\sigma}- \mbox{e}^{\lambda}\, \frac{
\dot{R}^2}{ R'^2} \right) F^2  \,, \label{9}\\
&&\nonumber\\
B(T,R) &=&
 \frac{  \mbox{e}^{\lambda} \, \mbox{e}^{2\sigma}}{
R'^2 \left( \mbox{e}^{2\sigma}- \mbox{e}^{\lambda}\, \frac{
\dot{R}^2}{ R'^2} \right)} \,.\label{10}
\end{eqnarray}
In his original paper, Lema\^itre identified (without 
a real justification) the mass contained in a sphere of radius $r$ 
with 
the quantity 
\be\label{11}
M= \frac{R}{2} \left( 1+\dot{R}^2 \mbox{e}^{-2\sigma} - 
R'^2 \mbox{e}^{-\lambda} -\frac{\Lambda R^2}{3} \right) \,.
\ee
The derivatives of this ``mass'' $M$ are related to the cosmic 
fluid density and pressure by \cite{Lemaitre, AH}
\begin{eqnarray}
M' &=& 4\pi R^2 R'\rho \,, \label{12}\\
&&\nonumber\\
\dot{M} &=& -4\pi R^2 \dot{R} P \,. \label{13}
\end{eqnarray}
In 1970, Cahill and McVittie \cite{CahillMcVittie} studied the 
Lema\^itre model with 
$\Lambda=0$ and identified the right hand side of eq.~(\ref{11}) 
(with $\Lambda=0$) with the physical mass. The rationale was that, 
if a Lema\^itre model is joined to an exterior Schwarschild 
geometry in a Swiss-cheese model, then $M$ must equal the exterior 
mass,\footnote{In 
retrospect, this is a good argument because there is little 
arguing 
on the physical mass of the Schwarzschild spacetime, and the choice 
proved to give the correct answer (see below).} and that the 
Bianchi identities generalize eqs.~(\ref{12}) and (\ref{13}). 
Cahill and McVittie were aware that their mass proposal $M$ 
coincided with the Misner-Sharp-Hernandez mass, then just recently 
introduced \cite{CahillMcVittie}.

The authors of Ref.~\cite{AH} note that the interpretation of $M$ 
is rather clear when there is no pressure and $\Lambda$ is absent,  
but it is not so straighforward otherwise. They proceed to note 
that eq.~(\ref{12}), which does not contain $P$, holds in all 
Lema\^itre models. This is a good observation but, in principle, 
there could be other 
quantities or equations which do not depend explicitly on $P$ and 
could be used to define effective masses. As a guideline to find 
out the 
physical mass of sphere, the authors of \cite{AH} proceed to 
analyze 
the geodesic deviation equation in order to establish which mass 
is ``seen'' by test particles, and the discussion necessarily 
becomes rather involved.  They reach the rather disheartening 
conclusion that ``we cannot separate the mass, the cosmological 
constant, the density and the pressure from each other, and so we 
cannot create a unique definition of mass based on geometric 
invariants of the metric in the general case'' \cite{AH}. Indeed, 
we can. Our knowledge of energy and mass in general relativity has 
progressed greatly since the times of Lema\^itre and Cahill and 
McVittie, with the introduction of the various quasi-local energies
(see \cite{Szabados} for a review), which culminated in the 
Hawking-Hayward quasi-local energy \cite{Hawking, Hayward1}. In 
spherical symmetry, the Hawking-Hayward quasi-local energy reduces 
\cite{Hayward1, Hayward2} to the Misner-Sharp-Hernandez mass 
$M_\text{MSH}$ \cite{MSH}, which is defined in a 
coordinate-invariant way by \cite{MSH, Hayward1, Hayward2, 
NielsenVisser, AbreuVisser, book}
\be
1-\frac{2M_\text{MSH}}{R}=\nabla^cR\nabla_c R \,,
\label{14}
\ee
where $R$ is the areal radius (which, being related to the area 
${\cal A}$ of 2-spheres of symmetry by $R=\sqrt{ \frac{{\cal 
A}}{4\pi} }$, is a geometrically defined quantity).

In coordinates $\left\{x^{\mu'} \right\}=\left\{  T, R, \theta, 
\varphi \right\}$, the squared 
gradient in the right hand side of eq.~(\ref{14}) is simply  
$g^{RR}=B^{-1}$ and gives the Misner-Sharp-Hernandez mass
\begin{eqnarray}
M_\text{MSH} &=& \frac{R}{2} \left( 1-g^{RR} \right) \nonumber\\
&&\nonumber\\
&=& \frac{R}{2} \left( 1+ \dot{R}^2 \mbox{e}^{-2\sigma} 
-R'^2 \, \mbox{e}^{-\lambda} \right) \nonumber\\
&&\nonumber\\
&=& M+\frac{\Lambda R^3}{6} \,. 
\label{15}
\end{eqnarray}
So, for $\Lambda=0$, the Cahill-McVittie prescription 
coincides 
with the Hawking-Hayward/Misner-Sharp-Hernandez one. This result 
can, of course, be 
obtained also in coordinates $\left\{ t,r,\theta, \varphi \right\} 
$ by computing $\nabla^cR\nabla_c R$ with the metric~(\ref{1}) and 
using the well known relation
\be
\dot{R}=\pm \mbox{e}^{\sigma} \sqrt{ \frac{2M}{R} + R'^2 \, 
\mbox{e}^{-\lambda} -1 +\frac{\Lambda R^2}{3} } \,,
\label{16}
\ee
which follows from the definition~(\ref{11}) \cite{AH}.

Note that the Misner-Sharp-Hernandez mass does not depend 
explicitly 
on the pressure $P$ (although $P$ affects the cosmic expansion and 
determines the metric coefficients which, in turn, determine 
$M_\text{MSH}$), while $\dot{M}$ depends on $P$ but not on $\rho$ 
(eq.~(\ref{13})). This fact is well known in FLRW space 
\cite{Hayward1, myPRDcosmohorizons}, to which the Lema\^itre model 
reduces if $\sigma \equiv 1 $ and $\lambda=\lambda(t)$. Therefore, 
there is no issue of disentangling the contribution of $P$ from 
those 
of $\rho$ and $\Lambda$. The contribution $\Lambda R^3/6$ to 
$M_\text{MSH}$ is easily interpreted as the mass corresponding to 
the 
cosmological constant energy density 
$\rho_{\Lambda}=\frac{\Lambda}{8\pi} $ contained in a sphere of 
areal radius $R$ and volume $4\pi R^3/3$. The decomposition of 
$M_\text{MSH}$ into a ``local'' and a ``cosmological''  part is 
covariant: this point has been discussed in detail in 
\cite{CarreraGiulini} for the McVittie metric \cite{McVittie}, 
which exhibits some 
of the properties of the Lema\^itre model (although it belongs to a 
different family of solutions of the Einstein equations), and we 
will not repeat the discussion here.

Although the Hawking-Hayward mass is not mentioned in \cite{AH}, 
the authors somehow end up reasoning along the same lines in their 
quest for the physical mass, when they stress the role of the 
apparent horizon in relating cosmic mass and diameter distance. 
In fact, the apparent horizons of any spherically symmetric metric 
are defined by $\nabla^c R \nabla_cR=0$ ({\em e.g.}, 
\cite{NielsenVisser, AbreuVisser}), which relates the apparent 
horizon radii with the Misner-Sharp-Hernandez mass 
contained through the relation  $R_\text{AH}=2M_\text{MSH}$ (which 
mimics
the expression of the Schwarzschild radius in the Schwarzschild 
spacetime) \cite{NielsenVisser, AbreuVisser}. In the Lema\^itre 
model, the recipe to locate the apparent horizons translates into 
$g^{RR}=0$, or
\be
R'^2 \mbox{e}^{2\sigma} - 
\dot{R}^2 \mbox{e}^{\lambda} =0 \,.
\label{17} 
\ee
In general, multiple solutions to this equation (which must be 
solved numerically) are possible, describing both black hole and 
cosmological apparent horizons which evolve in time (examples 
are solved in \cite{BoothBritsGonzalez, GaoetalLTB}). 

Using the Misner-Sharp-Hernandez mass~(\ref{15}), the line 
element~(\ref{8})  can now be written as
\be
ds^2=-\mbox{e}^{2\sigma+\lambda} \left( \frac{F}{R'} \right)^2 
\left( 1-\frac{2M_\text{MSH}}{R} \right)dT^2 
+\frac{dR^2}{ 1-\frac{2M_\text{MSH}}{R} }+R^2 d\Omega_{(2)}^2 \,.
\ee
The spatial part of this line element resembles the spatial part of 
the Schwarzschild metric, but only superficially because 
$M_\text{MSH}$ is not a constant but depends on the areal radius 
$R$.

\section{The Kodama energy current}
\label{sec:3}
\setcounter{equation}{0}

In a generic Lema\^itre model there is no timelike Killing vector 
but, as in any spherically symmetric spacetime, one can introduce 
the closest thing to it, the Kodama vector  \cite{Kodama}
\be\label{Kodamavector}
k^a=\epsilon^{ab}\nabla_b R \,,
\ee
where $\epsilon_{ab}$ is the volume element  on the 2-space 
orthogonal to the 2-spheres of symmetry. If the spacetime metric is 
decomposed according to
\be
ds^2=h_{ab} dx^a dx^b +R^2 d\Omega_{(2)}^2 \;\;\;\;\;\;\;\;
\;\;\;\;\; (a,b=0,1) \,,
\ee
then $\epsilon_{ab}$ is the volume form associated with the 
2-metric 
$h_{ab}$. The Kodama vector  
plays the role of a Killing vector where there is none: it is 
timelike in a spacetime 
region, null on an apparent horizon, and becomes spacelike on the 
other side of this horizon \cite{Kodama}. What makes the Kodama 
vector remarkable is the fact that the Kodama current 
\be\label{18}
J^a =G^{ab}k_b
\ee
associated with it is covariantly conserved \cite{Kodama}, 
$\nabla_c J^c=0$, such an unexpected  
property to be called the ``Kodama miracle'' \cite{AbreuVisser}. 
What is more, the Misner-Sharp-Hernandez mass almost universally 
identified with the physical mass-energy in spherical 
symmetry in 
general relativity, turns out to be the Noether charge associated 
with the Kodama current \cite{Hayward1}.  There are strong claims 
in the literature that, in the absence of a preferred time derived 
from a timelike Killing vector, the Kodama vector introduces a 
preferred time and surface gravity on apparent horizons, which 
determine a Hawking temperature and make thermodynamics meaningful 
for time-evolving {\em apparent} horizons (see 
Ref.~\cite{NadaliniVanzo, Galaxies2, book} for reviews). Given 
the geometry~(\ref{0}), the Kodama vector is \cite{Kodama}
\be\label{19}
k^a=\frac{-1}{\sqrt{AB}} \left( \frac{ \partial}{\partial T} 
\right)^a 
\ee
and, in the Lema\^itre geometry~(\ref{8}), it has components
\be\label{20}
k^{\mu'}=  -
\frac{|R'| \mbox{e}^{-\sigma}\mbox{e}^{-\lambda/2} }{F} \, 
\delta^{\mu ' 0} \,.
\ee

Cahill and McVittie found a conserved 4-current (reported 
also in \cite{AH}) which, in comoving coordinates $\left\{x^{\mu} 
\right\}= \left\{ t,r,\theta, \varphi \right\}$, has components
\be
J_\text{(CM)}^{\mu}= 
\frac{\sin\theta}{4\pi \sqrt{-g}} \left( M', -\dot{M}, 0,0 \right) 
= \frac{ 
\mbox{e}^{-\sigma}\mbox{e}^{-\lambda/2}}{4\pi R^2} 
\left( M', -\dot{M}, 0,0 \right) \,.
\ee
Eqs.~(\ref{12}) and (\ref{13}) then yield 
\be\label{24}
J_\text{(CM)}^{\mu} =   
\mbox{e}^{-\sigma}\mbox{e}^{-\lambda/2}
\left( R'\rho  , \dot{R}P , 0,0 \right) \,.
\ee
It is natural to ask whether this current is the same as the Kodama 
current. To find out, one computes the Kodama 
vector~(\ref{Kodamavector}), which is found to have components in 
comoving coordinates 
\be
k^{\mu}= \epsilon^{\mu\nu}\nabla_{\nu}R= \epsilon^{\mu 0} \dot{R}+ 
\epsilon^{\mu 1} R'\,.
\ee
Therefore, it is
\begin{eqnarray}
k^0 &=& \epsilon^{01}R'= g^{00}g^{11} \sqrt{|h|} \, 
R'=-\mbox{e}^{-\sigma} 
\mbox{e}^{-\lambda/2} R' \,,\\
&&\nonumber\\
k^1 &=& \epsilon^{10} \dot{R}= -g^{00}g^{11} \sqrt{|h|} \, 
\dot{R} = \mbox{e}^{-\sigma} \mbox{e}^{-\lambda/2} \dot{R} \,,
\end{eqnarray}
where $ h=-\mbox{e}^{2\sigma} \mbox{e}^{\lambda}$ is the 
determinant of the 2-metric $h_{ab}$ in the submanifold orthogonal 
to the 2-spheres of symmetry. We have
\be
k^{\mu}= \mbox{e}^{-\sigma} \mbox{e}^{-\lambda/2} 
\left( -R',  \dot{R}, 0,0 \right) 
\ee
and, lowering the indices,
\be
k_{\mu}= \left( \mbox{e}^{\sigma} \mbox{e}^{-\lambda/2} R',     
\mbox{e}^{-\sigma}\mbox{e}^{\lambda/2}  \dot{R}, 0,0 \right) \,. 
\ee
Using the Einstein equations~(\ref{efe}) and the fluid 
four-velocity $ u_{\mu}=-\mbox{e}^{\sigma} \, \delta_{\mu 0}$, the 
non-vanishing components of the Kodama current $J^a=G^{ab}k_b$ are 
found to be 
\begin{eqnarray}
J^0 &=& G^{00} k_0= (g^{00})^2 G_{00}k_0= 
\mbox{e}^{-\sigma} \, \mbox{e}^{-\lambda/2} R' 
\left( 8\pi \rho +\Lambda \right) \,,\\
&&\nonumber\\
J^1 &=& G^{11} k_1= (g^{11})^2 G_{11}k_1=
\mbox{e}^{-\sigma} \, \mbox{e}^{-\lambda/2} 
\dot{R}  \left( 8\pi P -\Lambda \right) \,,
\end{eqnarray} 
By comparing the expression
\be
J^{\mu}= \mbox{e}^{-\sigma} \, \mbox{e}^{-\lambda/2} \left( 
R' \left( 8\pi \rho +\Lambda \right) ,
\dot{R}  \left( 8\pi P -\Lambda \right) , 0, 0 \right)
\ee
with eq.~(\ref{24}), it is clear that the Cahill-McVittie conserved 
current is just 
\be
J^{\mu}_\text{(CM)}=(8\pi)^{-1}  J^{\mu}
\ee
when $\Lambda =0$ (and it is obtained by  contracting $T^{ab}$, 
instead  of 
$G^{ab}$, with the Kodama vector). This fact was to be suspected, 
since asking for two separate ``miracles'' would be asking too 
much.

\section{Conclusions}
\label{sec:5}
\setcounter{equation}{0}

There are several quasi-local energy constructs in general 
relativity (see the review \cite{Szabados}) and there is no 
mathematical ``proof'' selecting the ``correct'' one. However, 
there is now general consensus that the Hawking-Hayward quasi-local 
construction \cite{Hawking, Hayward1} is preferred in the sense 
that it encapsulates better than its competitors the physical 
properties required by the mass-energy of a system. Among the 
advantages of the Hawking-Hayward mass notion are the facts that it 
is 
well defined for non-asymptotically flat and non-stationary 
spacetimes \cite{Szabados}. In spherical 
symmetry, the Hawking-Hayward quasi-local energy reduces 
\cite{Hayward2}  to the better known 
Misner-Sharp-Hernandez mass \cite{MSH}. It appears that Cahill and 
McVittie identified the correct mass notion in the 
$\Lambda=0$ Lema\^itre 
space, the Hawking-Hayward/Misner-Sharp-Hernandez one. It is well 
known since the early days of the 
Hawking-Hayward mass \cite{Hayward1, Hayward2} that this object is  
also the Noether charge  associated with the Kodama current, and 
we won't repeat the derivation of this result here. The 
Cahill-McVittie covariantly conserved current is just a multiple of 
the Kodama energy current. Lema\^itre, Cahill, and 
McVittie made a clever choice and they were correct after all, 
although perhaps 
they could not be sure of the reason why, because they came decades 
before Hawking 
and Hayward or before the present-day understanding of the 
quasi-local energy.

\section*{Acknowledgments}
This work is supported by Bishop's University and by 
the Natural Sciences and Engineering Research Council of 
Canada ({\em NSERC}).

\end{document}